\documentclass[11pt]{article}
\usepackage{geometry}   
\usepackage{dsfont}
\usepackage{amsthm}
\usepackage{amsmath}
\usepackage{amssymb}
\usepackage{esint}
\usepackage{graphicx}
\usepackage{mathrsfs}
\usepackage{textpos}             % See geometry.pdf to learn the layout options. There are lots.
\usepackage{graphicx}
\usepackage{amssymb}
\usepackage{epstopdf}

\title{Entanglement Enabled Intensity Interferometry}
\author{Jordan Cotler$^1$ and Frank Wilczek$^{1,2}$\\
\small\it 1. Center for Theoretical Physics, MIT, Cambridge MA 02139 USA \\
\small \it 2. Origins Project, Arizona State University, Tempe AZ 25287 USA}

\begin{document}

\maketitle

\begin{textblock*}{5cm}(11cm,-8.2cm)
  \fbox{\footnotesize MIT-CTP-4641}
\end{textblock*}

%\section{}
%\subsection{}

\begin{abstract}
Intensity interferometry (Hanbury Brown - Twiss effect) is an interesting and useful concept that is usually presented as a manifestation of the quantum statistics of indistinguishable particles.  Here, by exploiting possibilities for projection and entanglement, we substantially widen the scope of its central idea, removing the requirement of indistinguishability.  We thereby potentially gain access to a host of new observables, including subtle polarization correlations and entanglement itself.   Our considerations also shed light on the physical significance of superselection.
\end{abstract}

Viewed from the perspective of quantum theory, interferometry is the exploitation of effects that arise when there are several ways to evolve from a given initial state into a given final state.   The probability for the total process is then the square of the sum of the transition amplitudes, which is not equal to the sum of the squares of the transition amplitudes.   Allowing for two alternatives, $A$, $B$, we have
\begin{equation}
| A + B|^2 ~=~ |A|^2 + |B|^2 + 2 {\rm Re } \, AB^*
\end{equation}
The difference, i.e.\!\! the interference term $ 2 {\rm Re } \, AB^*$, often encodes valuable information.   Since this term depends on the relative phase between $A$ and $B$, it will generally vanish for alternatives subject to independent phase noise.  Therefore, observation of interference typically requires coherent sources.

A particularly interesting, novel kind of interferometry was invented by Hanbury Brown and Twiss (HBT), which sidesteps that requirement \cite{HBT1}-\cite{Baym}.   In HBT ``intensity interferometry,'' the transition between initial state and final state involves emission and absorption of two indistinguishable photons at two locations (so the ``detector'' contains two distinct modules).  The overall result can occur through two distinct (crossed) channels, and by measuring the total rate one can study interference between those two channels.    

Hanbury Brown and Twiss, and many subsequent authors, emphasized the importance of identical particles and quantum statistics in enabling their effect.   Here we show that by allowing more complexity in the detectors, and in particular by exploiting projection and entanglement, one can relax those requirements.   The basic concept is that the same final state will be accessed from different {\it emission \/} patterns by way of different parts of the {\it detector \/} wave function.   One can arrange for the two differences to cancel one another.  We call this entanglement enabled intensity interferometry (E$^2$I$^2$), in general, although in some cases projection may substitute for entanglement.  

%With that inspiration we here investigate a strategy for interferometry, wherein one exploits the possibility of entanglement on the detector side.    Since the same final state may be accessed by processes that run through different parts of the detector wave function, such entanglement opens up new possibilities for interferometry.    

Entanglement enabled intensity interferometry allows one, in principle, to interfere sources that produce photons with orthogonal polarization, photons of different wavelengths, or photon and electron sources.   Entanglement enabled intensity interferometry demonstrates an important limitation on the concept of superselection, by exhibiting phenomena whose occurrence violates a local version of that concept.

Here we will supply a very general mathematical formulation of these ideas, consider several schematic designs for interferometers exploiting entanglement, discuss the insight it affords into the issue of superselection, and briefly indicate some potential applications.   Detailed designs and specific applications will appear in future work.

\bigskip

\section{Review of HBT}

We consider two sources of photons 1, 2 and two detectors $A, B$.  (See Figure 1.)   We denote the propagator from 1 to $A$ by $D_{1A}$, and similarly for the other source-detector combinations.   Neglecting also, for simplicity, the possible difference in arrival times, we find that the probability for simultaneous firing of both detectors due to emissions from the two sources involves 
\begin{equation}
| \, D_{1A}D_{2B} \, + \, D_{2A}D_{1B} \, |^2 ~=~  | D_{1A} |^2 \, | D_{2B}|^2 \, + \, | D_{2A}|^2 \, |D_{1B} |^2 \, + \, 2 \, {\rm Re} \, D_{1A}D_{2B}D_{2A}^*D_{1B}^*
\end{equation}
The first two terms in this expression are the contributions from the red and blue {\it processes\/} in Figure 1, while the third term represents interference between them.   This interference term has the remarkable property that random phases associated with the emitters 1 or 2 cancel, since each appears in both a propagator and a conjugate propagator.    The interference term therefore depends only on the relative phase factor introduced by the geometry of the situation.   As one varies the distance between the detectors, one gets positive or negative interference.  The distance between maxima reflects the separation of the sources.  For a single extended source, such as a star, the contrast will wash out at large detector separations.  The rate with which that happens reflects the angular size of the source, and can be used to measure it.  Hanbury Brown and Twiss exploited this effect to measure the radius of Sirius and of several other stars.   Subsequently, their basic idea has been applied in many other applications, ranging from heavy ion collisions to the study of condensed matter.  

${}$  \\
\begin{center}\label{hbtGeometry}
\includegraphics[scale=0.36]{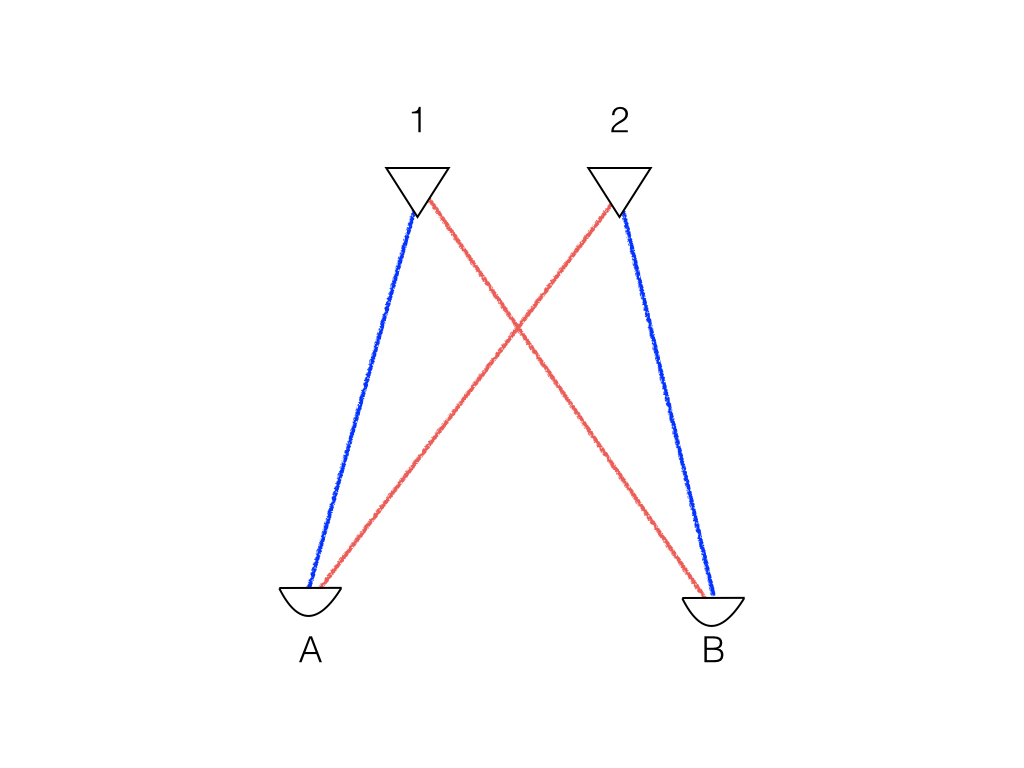}\\
Figure 1: Geometry of the Hanbury Brown - Twiss intensity interferometer.  Two distinct processes contribute to correlated firing of the detectors.
\end{center}
${}$  \\

\section{Polarization}

It was implicit, in our preceding discussion, that the detector could not reveal where its photon came from.
If the photons have orthogonal polarizations, for example, they will not interfere. 
For unpolarized sources, this effectively halves the HBT effect.  

The question naturally arises: If the emitters do have non-trivial polarization properties, can we access them? 
For example: If we have two very nearby sources, that emit in orthogonal polarizations, can we resolve them?  
Unadorned HBT will not serve here, but (as we shall see) a simple refinement accesses much more information, and does the job.  

Let us first consider the simple case where emitter 1 produces photons with polarization described, in a basis of orthogonal linear polarizations, by 
$\left(\begin{array}{c}\alpha \\ \beta \end{array}\right)$, while emitter 2 produces photons with polarization $\left(\begin{array}{c}\gamma \\ \delta\end{array}\right)$.  Furthermore, let us apply projections $\Pi_A, \Pi_B$ at the two detectors.   Then the rate for simultaneous firing becomes, as in our earlier discussion, a sum of the separate process terms
\begin{eqnarray}
&{}&\left(\begin{array}{cc}\alpha^* & \beta^*\end{array}\right) \Pi_A   \left(\begin{array}{c}\alpha \\\beta\end{array}\right)  \ \,
\left(\begin{array}{cc}\gamma^* & \delta^*\end{array}\right)  \Pi_B  \left(\begin{array}{c}\gamma \\\delta\end{array}\right)\  |D_{1A}|^2 |D_{2B}|^2 \ +\nonumber \\
&{}&
\left(\begin{array}{cc}\gamma^* & \delta^*\end{array}\right)  \Pi_A  \left(\begin{array}{c}\gamma \\\delta\end{array}\right)  \ \, 
\left(\begin{array}{cc}\alpha^* & \beta^*\end{array}\right)  \Pi_B  \left(\begin{array}{c}\alpha \\\beta\end{array}\right) \ |D_{2A}|^2 |D_{1B}|^2 
\end{eqnarray}
and the interference term
\begin{equation}
\left(\begin{array}{cc}\gamma^* & \delta^*\end{array}\right)  \Pi_A  \left(\begin{array}{c}\alpha \\\beta\end{array}\right)  \ \, 
\left(\begin{array}{cc}\alpha^* & \beta^*\end{array}\right)  \Pi_B  \left(\begin{array}{c}\gamma \\\delta\end{array}\right)\ \, 
D_{1A}D_{2B}D_{2A}^*D_{1B}^* \ + \ {\rm c.c.}
\end{equation}

We can generalize this by allowing the sources to emit in mixtures, described by polarization (density) matrices $\pi_1, \pi_2$.  Then we get for the uncrossed terms
\begin{equation}\label{direct}
{\rm Tr} \,  \Pi_A  \pi_1  \, {\rm Tr} \,  \Pi_B  \pi_2 \ |D_{1A}|^2 |D_{2B}|^2 \, + \,
{\rm Tr} \,  \Pi_A  \pi_2  \, {\rm Tr} \,  \Pi_B  \pi_1 \  |D_{2A}|^2 |D_{1B}|^2 
\end{equation}
and for the crossed term
\begin{eqnarray}\label{crossed}
&{}& {\rm Tr} \, \Pi_A \pi_1 \Pi_B \pi_2 \ D_{1A}D_{2B}D_{2A}^*D_{1B}^* \, + \, {\rm c.c.}  \\
&=& {\rm Tr} \, \Pi_A \pi_1 \Pi_B \pi_2 \ D_{1A}D_{2B}D_{2A}^*D_{1B}^* \, + \, {\rm Tr} \, \Pi_A \pi_2 \Pi_B \pi_1 \ D_{1A}^*D_{2B}^*D_{2A}D_{1B} \nonumber
\end{eqnarray}
where in the second line we exploit the hermiticity of $\pi, \Pi$.   

By letting the $\Pi$s interpolate between the two orthogonal polarizations of the sources in our model problem, we obtain interference between them, which could allow us to resolve them.   More generally, use of the $\Pi$s can significantly enhance our perception of the sources.   

With $\Pi_A = \Pi_B = 1$ intensity interferometry accesses the {\it cross-polarization}, an interesting quantity that has been discussed previously \cite{Wolf1}-\cite{Wolf3}.    We will call the more general phenomenon linked polarization.   

Note that in this procedure we have {\it gained \/} one form of information by {\it erasing\/} potential information that would have enabled us, in principle, to say which source was responsible for the emission (even when we can't resolve it spatially).   That erasure renders two otherwise distinguishable processes to become indistinguishable, and enables their interference. 

\section{Introducing Entanglement}

So far, we have used selective projection to get interference between non-identical emissions.   A more general and powerful technique also exploits entanglement of the detectors.  As an extreme example, let us consider that one of our emitters emits bosons $b$, while the other emits fermions $f$. 
A detector that receives a boson goes into state {\bf B}, while a detector that receives a fermion goes into state {\bf F}.  (See Figure 2.)  

${}$  \\
\begin{center}\label{hbtDistinguishable}
\includegraphics[scale=0.36]{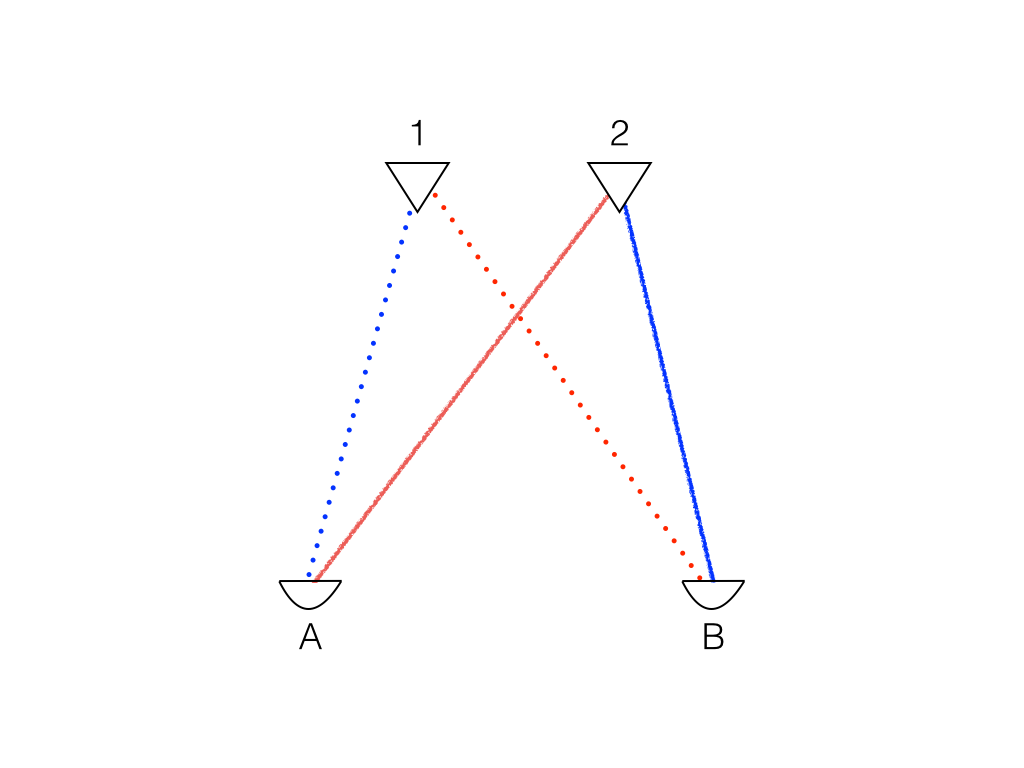}\\
Figure 2: The same geometry as in Figure 1, but drawn to emphasize the possibility of distinguishable emissions. 
\end{center}
${}$  \\

We would like to get interference between the terms in 
\begin{equation*}
S_{1A}\, D_{2B} | \, {\bf FB} \rangle \, + \, D_{2A} \, S_{1B} \, | {\bf BF} \rangle
\end{equation*}
Following a similar philosophy to our polarization example, we change the state basis and erase information to access interference.

We can do that directly, using entangled detector states (Procedure 1).  Writing 
\begin{eqnarray}
&{}& S_{1A}\, D_{2B} | \, {\bf FB} \rangle \, + \, D_{2A} \, S_{1B} \, | {\bf BF} \rangle \nonumber \\
&=& \frac{1}{2} ( S_{1A}\, D_{2B} \, + \, D_{2A} \, S_{1B} ) (| \, {\bf FB} \rangle \, + \, | {\bf BF} \rangle) \nonumber \\
&+& \frac{1}{2} ( S_{1A}\, D_{2B} \, - \, D_{2A} \, S_{1B} ) (| \, {\bf FB} \rangle \, - \, | {\bf BF} \rangle) \nonumber 
\end{eqnarray}
we see that by projecting on the entangled state
\begin{equation*}
\frac{1}{\sqrt 2} ( |\,{\bf FB} \rangle \, + \, | {\bf BF} \rangle )
\end{equation*}
we measure
\begin{equation*}
| S_{1A} D_{2B} \, + \, D_{2A} S_{1A} |^2
\end{equation*}

We might also follow the polarization strategy more literally, acting on the detectors separately (Procedure 2).  Here, with 
\begin{eqnarray}
| {\bf F} \rangle_A ~&=&~ \frac{1}{\sqrt 2} ( | C \rangle \, + \, | D \rangle) \nonumber \\
| {\bf B} \rangle_A ~&=&~ \frac{1}{\sqrt 2} ( | C \rangle \, - \, | D \rangle) \nonumber \\
| {\bf F} \rangle_B ~&=&~ \frac{1}{\sqrt 2} ( | E \rangle \, + \, | F \rangle) \nonumber \\
| {\bf B} \rangle_B ~&=&~ \frac{1}{\sqrt 2} ( | E \rangle \, - \, | F \rangle) 
\end{eqnarray}
projection on 
\begin{equation*}
| C \rangle \, \langle C | \, \otimes | E \rangle \, \langle E |
\end{equation*}
gives us what we want.  

\section{Superselection}

Procedure 2 requires that we set up coherent superpositions of states that differ by one in fermion number.   That may be problematic, because it violates a superselection rule.   Procedure 1 is free of that issue, because the two parts of $\frac{1}{\sqrt 2} ( |\, {\bf FB} \rangle \, + \, | {\bf BF} \rangle )$, while they differ in fermion number locally, agree in that respect globally.   Below, we shall indicate a geometrical method for realizing this entanglement.   The possibility of measurable boson-fermion interference sheds an interesting light on superselection, emphasizing its global nature.  

\section{General Entanglement}

We can capture both our procedures, and reach a proper generalization, in the following way.   The projectors $\Pi_A, \Pi_B$ encode density matrices for the final states of the detectors $A, B$.  When their states are entangled, however, the density matrix of the entire system will not factorize, and we will need to employ a system density matrix, in the form
\begin{equation*}
(\Pi_A)^{\alpha_1}_{\alpha_2} \, (\Pi_B)^{\beta_1}_{\beta_2} \, \rightarrow \, {\bf \Pi}^{\alpha_1 \beta_1}_{\alpha_2 \beta_2}
\end{equation*}

We should also allow for the interesting possibility of entanglement in the emitters.   That is accommodated according to 
\begin{equation*}
(\pi_1)^{\alpha_1}_{\alpha_2} \, (\pi_2)^{\beta_1}_{\beta_2} \, \rightarrow \, {\boldsymbol \pi}^{\alpha_1 \beta_1}_{\alpha_2 \beta_2}
\end{equation*}

With these notations, we can generalize our formula Eqn.\,(\ref{direct}) in the form
\begin{equation}\label{genDirect}
{\bf \Pi}^{\alpha_1 \beta_1}_{\alpha_2 \beta_2} \, {\boldsymbol \pi}^{\alpha_2 \beta_2}_{\alpha_1 \beta_1} \, |D_{1A}|^2 |D_{2B}|^2 \, + \,
 {\bf \Pi}^{\alpha_1 \beta_1}_{\alpha_2 \beta_2} \, {\boldsymbol \pi}^{\beta_2 \alpha_2}_{\beta_1 \alpha_1} \,  |D_{2A}|^2 |D_{1B}|^2 
 \end{equation}
and our formula Eqn.\,(\ref{crossed}) in the form
\begin{equation}\label{genCrossed}
{\bf \Pi}^{\alpha_1\beta_1}_{\alpha_2 \beta_2} \, {\boldsymbol \pi}_{\alpha_1 \beta_1}^{\beta_2 \alpha_2}  \ D_{1A}D_{2B}D_{2A}^*D_{1B}^* \, 
+ \, 
{\bf \Pi}^{\alpha_1\beta_1}_{\alpha_2 \beta_2} \, {\boldsymbol \pi}_{\beta_1 \alpha_1}^{\alpha_2 \beta_2} \ D_{1A}^*D_{2B}^*D_{2A}D_{1B}
\end{equation}

By comparing experimental data with Eqns.\,(\ref{genDirect}, \ref{genCrossed}), and determining whether it is consistent with factorization of ${\boldsymbol \pi}$, we become sensitive to entanglement between the emitters.   This effect could be used as a probe for proposed exotic states of matter that feature long-range entanglement.   It would also be interesting to investigate the possible existence of entanglement and linked polarization in the microwave background radiation.   

% Study of fluorescence emission by inhomogeneous materials could be used to characterize them.

\section{Variations}

%\subsection{Orthogonal Channels}

\subsection{Entanglement by Spatial Superposition}
Previously, in order to measure interference between the states of a boson and fermion, we needed to project detector states onto $\frac{1}{
\sqrt{2}}(|\,\textbf{FB}\rangle + |\textbf{BF}\rangle)$.  Here we describe a more geometric way to obtain the desired interference.

Say that we have two detectors located at $A$ and $B$ respectively.  The detector located at $A$ begins in a fermion accepting state $|\mathbb{F}\rangle$ which transitions to $|\textbf{F}\rangle$ if and only if it absorbs a fermion.  Similarly, the detector at $B$ begins in a boson accepting state $|\mathbb{B}\rangle$ which transitions to $|\textbf{B}\rangle$ if and only if it absorbs a boson.  The initial state of the detectors is $|\mathbb{F}\rangle|A\rangle \otimes |\mathbb{B}\rangle|B\rangle$ where we have included states that keep track of the positions of the detectors.

Now we put the detectors in an equal superposition of being at their original positions and being in swapped positions as
\begin{equation} \label{detectorSwap}
\frac{1}{\sqrt{2}} \left(|\mathbb{F}\rangle|A\rangle \otimes |\mathbb{B}\rangle|B\rangle + |\mathbb{F}\rangle|B\rangle \otimes |\mathbb{B}\rangle|A\rangle \right)
\end{equation}
This superposition of swapped and unswapped detectors can be obtained by application of the unitary operator
\begin{align}
\widehat{S} = &\frac{1}{\sqrt{2}} \left( |A\rangle \otimes |B\rangle +  |B\rangle \otimes |A\rangle \right) \left(\langle A| \otimes  \langle B| \right) + \frac{1}{\sqrt{2}} \left( |A\rangle \otimes |B\rangle -  |B\rangle \otimes |A\rangle \right) \left(\langle B| \otimes  \langle A| \right) \nonumber \\
&+ |A\rangle\langle A| \otimes |A\rangle\langle A| + |B\rangle\langle B| \otimes |B\rangle\langle B| \nonumber
\end{align}
to the spatial states.  Note that $\widehat{S}$ has the property $\widehat{S}^2 = \textbf{1}$.

If we have a fermion emitter and a boson emitter as before, we will be interested in the terms
\begin{equation} \label{detectorSwapRegistered}
S_{1A} D_{2B} |\textbf{F}\rangle |A\rangle \otimes |\textbf{B}\rangle |B\rangle + D_{2A} S_{1B}  |\textbf{F}\rangle |B\rangle \otimes |\textbf{B}\rangle |A\rangle
\end{equation}
Applying $\widehat{S}$ to the terms in Eqn. (\ref{detectorSwapRegistered}) we obtain
\begin{equation}
\frac{1}{\sqrt{2}} ( S_{1A} D_{2B} + D_{2A} S_{1B} ) |\textbf{F}\rangle |A\rangle \otimes |\textbf{B}\rangle |B\rangle + \frac{1}{\sqrt{2}} ( S_{1A} D_{2B} - D_{2A} S_{1B} ) |\textbf{F}\rangle |B\rangle \otimes |\textbf{B}\rangle |A\rangle\nonumber
\end{equation}
Projecting on the separable state $ |\textbf{F}\rangle |A\rangle \otimes |\textbf{B}\rangle |B\rangle$, we measure $|S_{1A}D_{2B} + D_{2A}S_{1B}|^2$ as desired.

In practice, of course, maintaining quantum coherence through a spatial swap operation presents a demanding challenge, but at least it is a clearly defined one.

\subsection{Different Wavelengths}

The preceding strategy, which enables our Procedure 1, might also be applied to the interference of photons with different wavelengths.   It is also not difficult to devise, for this application, a quasi-realistic version of Procedure 2, with projection at the separate detectors.   (See Figure 3.)   We imagine producing a definite superposition of atomic states $\alpha | 0 \rangle + \beta |1 \rangle$, both of which can transition, upon reception of different-wavelength photons, to a common level $| 2 \rangle$.  Then the rate for populating $| 2 \rangle$ will depend on the relative phase between those photons, and exhibits their interference.  
 
${}$  \\
\begin{center}\label{atomicLevels}
\includegraphics[scale=0.5]{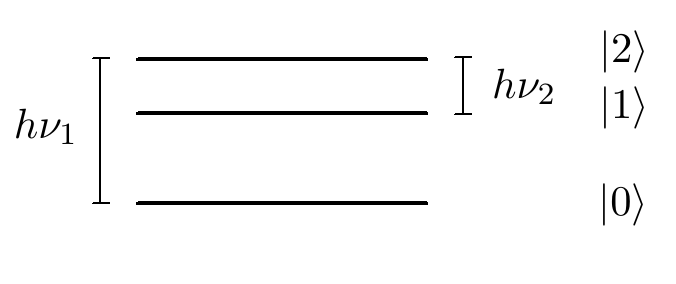}\\
Figure 3: States of an atom, showing two levels that may both feed into a third.  
\end{center}
${}$  \\ 

\subsection{Single Source of Decaying Particles}
A central theme of E$^2$I$^2$ is measuring detector states in mixed bases.  We can leverage this technique in situations when there is only a single source.  Here we will discuss the interesting case of a single source that is a collection of identical decaying particles.

Consider a particle $C$ which either decays into two particles of type $D$ or two particles of type $E$.  Thus, the two decay channels are $C \to DD$ and $C \to EE$ which occur with probability amplitudes $\mathcal{M}_{C \to DD}$ and $\mathcal{M}_{C \to EE}$ respectively.  Typically, when we compute information about a decay process we are interested in the absolute squares of the probability amplitudes $|\mathcal{M}_{C \to DD}|^2$, $|\mathcal{M}_{C \to EE}|^2$ which are measured in standard particle experiments.  By considering mixed bases of detector states, we gain access to the relative phases between the probability amplitudes.

Let ``$1$" denote a source of decaying $C$ particles.   Additionally, we have two detectors at $A$ and $B$ which are in $DD$ and $EE$ accepting states respectively.  We want interference between the terms
\begin{equation}
\mathcal{M}_{C \to DD} D_{1A} |\textbf{DD}\rangle |\mathbb{EE}\rangle + \mathcal{M}_{C \to EE} S_{1B} |\mathbb{DD}\rangle |\textbf{EE}\rangle  \nonumber
\end{equation}
Projecting onto $\frac{1}{\sqrt{2}} (|\textbf{DD}\rangle |\mathbb{EE}\rangle + |\mathbb{DD}\rangle |\textbf{EE}\rangle) $ we obtain
\begin{equation}
|\mathcal{M}_{C \to DD} D_{1A} + \mathcal{M}_{C \to EE} S_{1B}|^2 \nonumber
\end{equation}
which gives interference between the probability amplitudes for decay.  A similar procedure can be used to measure relative phases between the probability amplitudes of scattering processes.

\subsection{Mach-Zehnder Interferometry without Recombining Paths}

The basic concept of E$^2$I$^2$ is very flexible.  As an example of the possibilities, we outline an alternative realization of the classic Mach-Zehnder interferometer, wherein entanglement allows us to simplify the geometry. 

In Mach-Zehnder interferometry, two alternative paths for a single photon are produced by interposing a beamsplitter and recombining at a detector.  One can complicate propagation in one of the paths by inserting a translucent material and thereby, through the modified interference, obtain information about the material.  Using the techniques of E$^2$I$^2$, we can measure Mach-Zehnder interference \textit{without}  recombination of paths.  Consider the setup in Figure 4.
${}$  \\
\begin{center}
\label{NonlocalMZ}
\includegraphics[scale=0.36]{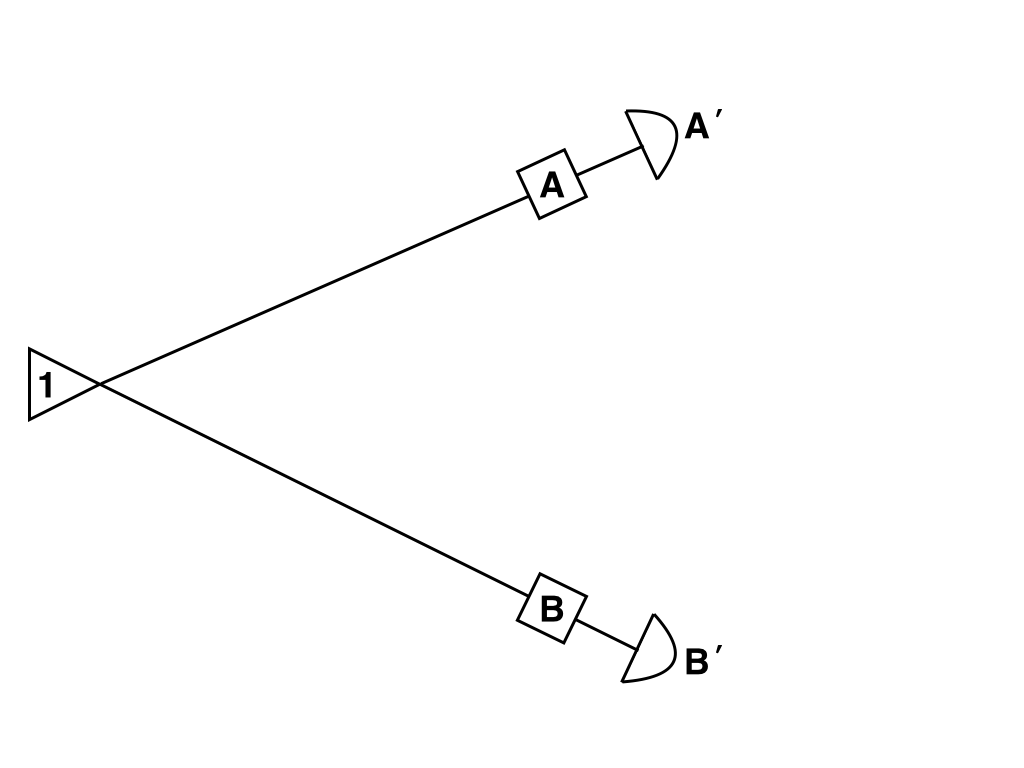}\\
$${}$$
Figure 4: Geometry of setup to perform Mach-Zehnder interferometry without recombining paths.
\end{center}
${}$  \\
We have a source (labeled ``$1$") of photons of energy $h \nu$, two boxes which each contains atoms of type $A$ and $B$ respectively, each in the state $\frac{1}{\sqrt{2}}(|0\rangle + |1\rangle)$ (see Figure 5), and two detectors $A'$ and $B'$.
${}$  \\
\begin{center}
\label{moreAtomicLevels}
\includegraphics[scale=0.5]{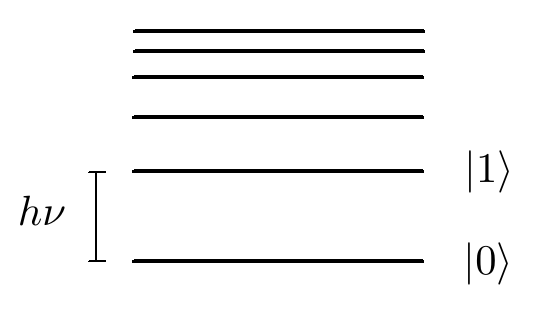}\\
Figure 5: States of the atoms $A$ and $B$.
\end{center}
${}$  \\
\indent If a photon from the source interacts with an atom in the ground state $|0\rangle$, the atom will be excited to $|1\rangle$ and the detector behind the atom will \textit{not} fire.  On the other hand, if a photon from the source interacts with an atom in the excited state $|1\rangle$, then the atom will not absorb the photon and instead the photon will be absorbed by the nearby detector.  We want to consider the terms
\begin{eqnarray*}
&{}& D_{1A} |1\rangle_A \otimes \frac{1}{\sqrt{2}}(|0\rangle_B + |1\rangle_B) \otimes |\text{not fired}\rangle_{A'} \otimes |\text{not fired}\rangle_{B'} \\
&+& D_{1B} \frac{1}{\sqrt{2}}(|0\rangle_A + |1\rangle_A)\otimes |1\rangle_B \otimes |\text{not fired}\rangle_{A'} \otimes |\text{not fired}\rangle_{B'}
\end{eqnarray*}
Projecting onto $|1\rangle_A \otimes |\text{not fired}\rangle_{A'}$ and $|1\rangle_B \otimes |\text{not fired}\rangle_{B'}$, each which can be done locally, we get (up to a constant factor)
$$|D_{1A} + D_{1B}|^2$$
which contains the Mach-Zehnder interference term.  Note that the $A$, $A'$ and $B$, $B'$ arrangements can be far apart, but Mach-Zehnder type interference will occur nevertheless.

\subsection*{Acknowledgements}
Jordan Cotler was funded by the Undergraduate Research Opportunities Program at the Massachusetts Institute of Technology.  Frank Wilczek's work is supported by the U.S. Department of Energy under grant Contract Number  DE-SC00012567.

%\section{Conclusion}
%E$^2$I$^2$ is a new tool to enhance one's perception of the quantum world by allowing access to previously inaccessible observables.  Our new and novel form of interferometry leverages the quantum nature of detectors to interfere distinguishable particles.  The proposed model designs are within the realm of experimental possibility.

\end{document}